\documentclass[apjl,twocolappendix]{emulateapj}

\usepackage{acronym}
\usepackage{graphicx}
\usepackage{amsmath,amsfonts,amssymb}
\usepackage{mathrsfs}
\usepackage[utf8]{inputenc}
\usepackage[normalem]{ulem}

\usepackage[plainpages=false, colorlinks=true, anchorcolor=blue, linkcolor=blue, citecolor=blue, bookmarks=false]{hyperref}

\newcommand{\ie}{\textit{i.e.}}
\newcommand{\eg}{\textit{e.g.}}

\usepackage{color}

\begin{document}

\title{Probing Extreme-Density Matter with Gravitational Wave
Observations of\\ Binary Neutron Star Merger Remnants}

\author{David \surname{Radice}}
\affiliation{Institute for Advanced Study,
1 Einstein Drive, Princeton, NJ 08540, USA}
\affiliation{Department of Astrophysical Sciences, Princeton University,
4 Ivy Lane, Princeton, NJ 08544, USA}
\author{Sebastiano \surname{Bernuzzi}}
\affiliation{Department of Mathematical, Physical and Computer Sciences, University of Parma, I-43124 Parma, Italy}  
\affiliation{Istituto Nazionale di Fisica Nucleare, Sezione Milano Bicocca, gruppo collegato di Parma, I-43124 Parma, Italy}  
\author{Walter \surname{Del Pozzo}}
\affiliation{Dipartimento di Fisica ``Enrico Fermi'', Universit\`a di Pisa, Pisa I-56127, Italy }
\author{Luke F. \surname{Roberts}}
\affiliation{NSCL/FRIB and Department of Physics \& Astronomy, Michigan State University, 640 S Shaw Lane East
Lansing, MI 48824, USA}
\author{Christian D. \surname{Ott}}
\affiliation{TAPIR, Walter Burke Institute for Theoretical Physics,
  California Institute of Technology, 1200 E
  California Blvd, Pasadena, CA 91125, USA}


\begin{abstract}
We present a proof-of-concept study, based on numerical-relativity
simulations, of how gravitational waves (GWs) from neutron star merger
remnants can probe the nature of matter at extreme densities. Phase
transitions and extra degrees of freedom can emerge at densities beyond
those reached during the inspiral, and typically result in a softening
of the equation of state (EOS). We show that such physical effects
change the qualitative dynamics of the remnant evolution, but they are
not identifiable as a signature in the GW frequency, with the
exception of possible black-hole formation effects. The EOS softening
is, instead, encoded in the GW luminosity and phase and is in principle
detectable up to distances of the order of several Mpcs with advanced
detectors and up to hundreds of Mpcs with third generation detectors.
Probing extreme-density matter will require going beyond the current
paradigm and developing a more holistic strategy for modeling and
analyzing postmerger GW signals.
\end{abstract}


\section{Introduction}

Gravitational waves (GWs) \acused{GW} from merging \acp{NS} offer a
unique way to probe the physics of matter at densities a few times that
of nuclear saturation.  The phase evolution of the \ac{GW} signal in the
last several orbits before contact is affected by the stars' response to
the companion tidal field. Its measurement could provide a model
independent way to infer the Love number of the \acp{NS} and thus the
\ac{NS} radii with a precision of ${\sim} 1\, \mathrm{km}$ or
better~\citep{damour:2012yf, read:2013zra, delpozzo:2013ala,
bernuzzi:2014owa, hotokezaka:2016bzh, hinderer:2016eia, lackey:2016krb}.
Given that the mass distribution of known binary \acp{NS} is sharply
peaked around $1.35\, M_\odot$ \citep{lattimer:2012nd}, the inspiral
phase will not probe the properties of matter at the highest densities
that can be reached in \acp{NS}, which can have masses up to at least
$\sim 2\, M_\odot$ \citep{demorest:2010bx, antoniadis:2013pzd}.

\begin{figure*}
  
  \begin{minipage}{0.329\textwidth}
    \includegraphics[width=\textwidth]{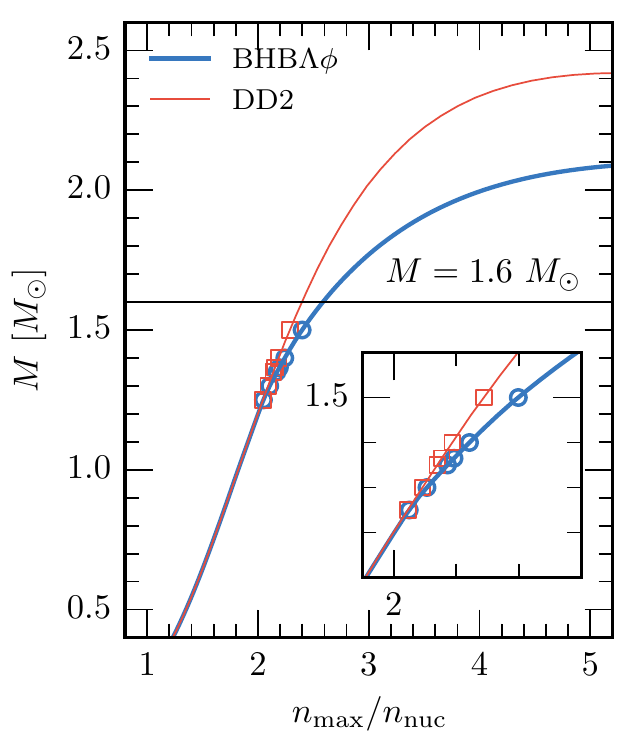}
  \end{minipage}
  \hfill
  \begin{minipage}{0.658\textwidth}
    \includegraphics[width=\textwidth]{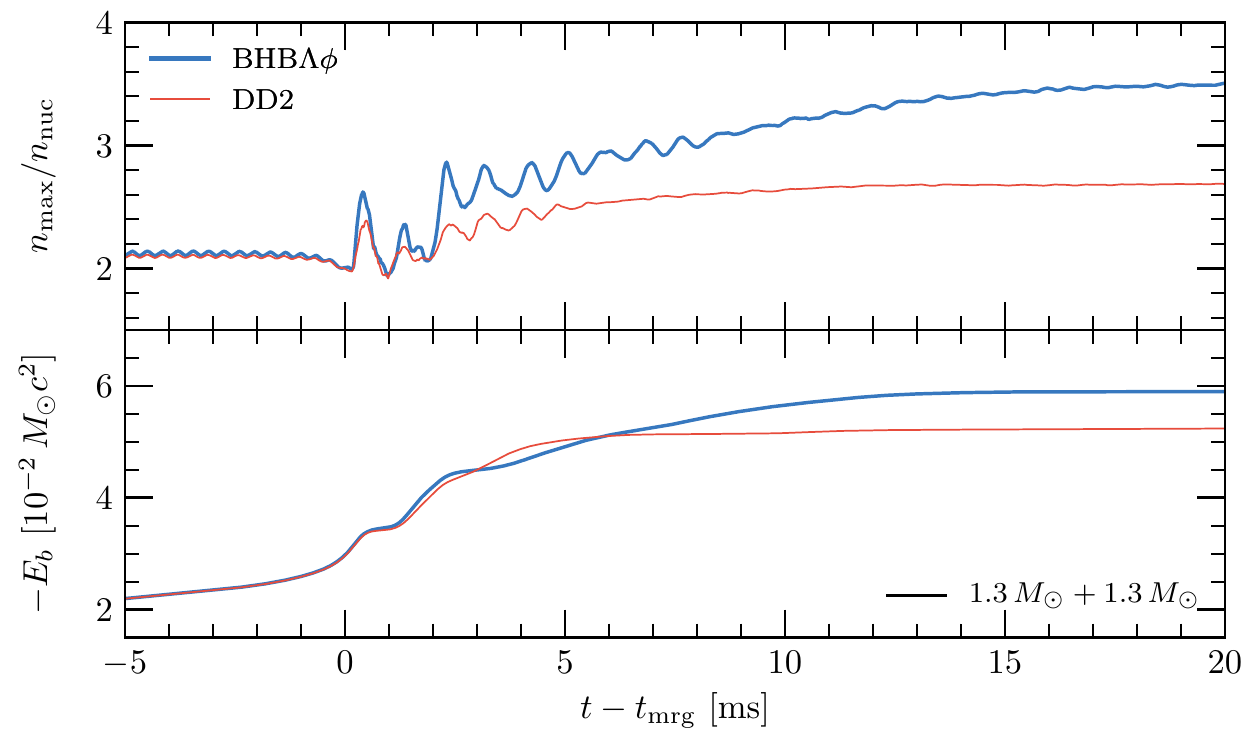}
  \end{minipage}
  \caption{\emph{Left panel:} mass-central density relations for
  spherical isolated \acp{NS} constructed with the BHB$\Lambda\phi$ and
  DD2 \ac{EOS}. The symbols denote the individual components of the
  binaries we consider here. \emph{Right panel:} maximum density
  (\emph{top}) and binding energy (\emph{bottom}) of the merger remnant
  relative to the binary at infinite separation for representative
  equal-masses binaries. The merger remnant becomes more compact and
  more bound with the BHB$\Lambda\phi$ \ac{EOS}. The central densities
  reach values comparable to those in the most massive isolated
  \acp{NS}.}
  \label{fig:compactness}
\end{figure*}

On the other hand, it is expected that the most common outcome of
\ac{NS} mergers is the formation of a compact remnant temporarily
supported against gravitational collapse by (differential) rotation over
timescales of several milliseconds to minutes after
merger~\citep{rosswog:2001fh, shibata:2006nm, baiotti:2008ra,
sekiguchi:2011zd, palenzuela:2015dqa, foucart:2015gaa, baiotti:2016qnr}.
This remnant is an efficient emitter of
\acp{GW}~\citep{bernuzzi:2015opx}. Their spectrum is complex and its
most prominent feature is a broad peak at frequency $f_2\sim 2-4\,
\mathrm{kHz}$ \citep{bauswein:2011tp, stergioulas:2011gd,
takami:2014zpa, bauswein:2015yca, rezzolla:2016nxn, dietrich:2016hky}.

For a fixed total binary mass, there is an empirical relation linking
$f_2$ and $R_{1.6}$, the radius of an isolated non-rotating $1.6\,
M_\odot$ \ac{NS} \citep{bauswein:2011tp, hotokezaka:2013iia}.  It has
also been argued that the behavior of the derivative of $f_2$ with
respect to the total mass could be used to constrain the maximum \ac{NS}
mass \citep{bauswein:2014qla}. The existence of these relations suggests
that $f_2$ can be used to infer the properties of matter at densities
larger than those achieved in the inspiral.  \citet{bernuzzi:2015rla},
however, showed the existence of a universal relation between $f_2$ and
the binary tidal coupling constant $\kappa_2^T$
\citep[\eg,][]{bernuzzi:2014kca}, which is a quantity encoding the
\emph{inspiral} properties of the binaries. Similar universal relations
have also been found for other characteristic frequencies of the signal
\citep{rezzolla:2016nxn}.

Is it then possible to probe the \ac{EOS} at the highest densities with
\ac{GW} observations? For example, could \ac{GW} observations of a
\ac{NS} merger remnant identify phase transitions occurring at densities
larger than those of the inspiral, but still relevant for massive
isolated \acp{NS}?  In this work, we show that the answer to both
questions is ``yes''. However, these measurements are not possible on
solely on the basis of existing fits for $f_2$ frequency in the
postmerger \ac{GW} spectrum.

Instead, they will require more sophisticated waveform modeling and
analysis.

\section{Methods} %

As a case study, we consider the late inspiral and merger of \ac{NS}
binaries simulated in full general relativity. We adopt two temperature
and composition-dependent \ac{EOS} for this work: the DD2 \ac{EOS}
\citep{typel:2009sy, hempel:2009mc} and the BHB$\Lambda\phi$ \ac{EOS}
\citep{banik:2014qja}. Both use the same description of nuclear matter,
but the BHB$\Lambda\phi$ \ac{EOS} also includes $\Lambda-$hyperons,
self-interacting via $\phi-$meson exchange.  Both \acp{EOS} are
consistent with theoretical and experimental constraints and with
astronomical observations of massive \acp{NS}. The DD2 and
BHB$\Lambda\phi$ \ac{EOS} predict maximum \ac{NS} masses of $2.42\,
M_\odot$ and $2.11\, M_\odot$, and $R_{1.6} = 13.17\, {\rm km}$ and
$R_{1.6} = 13.27\, \mathrm{km}$, respectively.  The difference in
$R_{1.6}$ is within the nominal error bars of the fits by
\citet{bauswein:2011tp} and \citet{hotokezaka:2013iia}, so they both
predict very similar $f_2$ GW frequency. We show the mass -- central
density curves for both \acp{EOS} in the left panel of
Fig.~\ref{fig:compactness}; see Fig.~2 of \citet{banik:2014qja} for the
mass -- radius curves. Although our quantitative results are specific to
these two \acp{EOS}, we expect our conclusions to generalize to all
\acp{EOS} for which a high density phase transition would be allowed by
current constraints, in particular the existence of $\simeq 2\ M_\odot$
\acp{NS}. The reason being that the main effects are a consequence of
the \ac{EOS} softening at densities larger than $\simeq 2.2\, n_{\rm
nuc}$ and are not specific to the appearance of $\Lambda$-particles in
the BHB$\Lambda\phi$ \ac{EOS}. Here, we take $n_{\rm nuc} = 0.16\, {\rm
fm}^{-3}$ (\ie, $\rho_{\rm nuc} \simeq 2.7\times 10^{14}\, {\rm g}\ {\rm
cm}^{-3}$) as the nuclear saturation density.

We consider $7$ binaries with total (isolation) masses between $2.5\,
M_\odot$ and $3\, M_\odot$, including $2$ unequal-mass cases. We evolve
each binary using both \acp{EOS}. For clarity, we discuss our
qualitative results using three representative equal-mass binaries.
Isolation \ac{NS} masses and central densities of these binaries are
highlighted in the left panel of Fig.~\ref{fig:compactness} and a full
list is given in Fig.~\ref{fig:logBayes}. We simulate the last ${\sim}
3$ orbits before merger and the evolution up to $21\, {\rm ms}$ after
the time of merger $t_{\rm mrg}$, defined as the peak time of the
\ac{GW} strain amplitude.

\begin{figure*}
  \includegraphics[width=\textwidth]{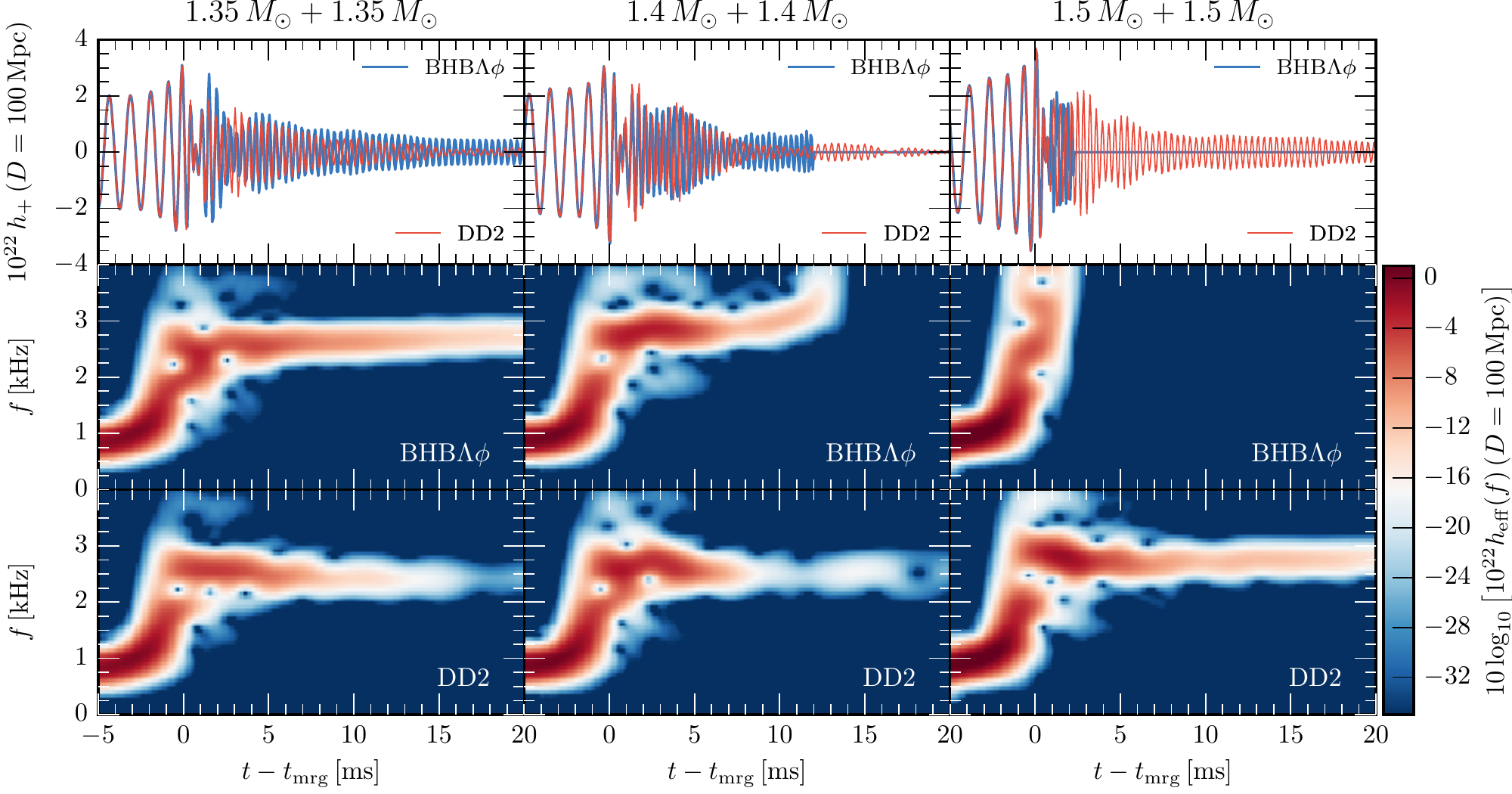}
  \caption{\ac{GW} strain (\emph{top panel}) and spectrograms
  (\emph{bottom panels}) for three representative binaries assuming
  optimal orientation for the ``$+$'' polarization. The appearance of
  hyperons enhances the \ac{GW} luminosity, but has only a modest impact
  on the \ac{GW} peak frequency. The latter does not shows a significant
  evolution until shortly before \ac{BH} formation.}
  \label{fig:waves.spec}
\end{figure*}

For the simulations, we use the \texttt{WhiskyTHC} code
\citep{radice:2013xpa}, with the high-resolution setup described in
\citet{bernuzzi:2015opx}, improved with conservative mesh-refinement,
not assuming rotational symmetry for equal mass binaries, and extracting
\acp{GW} at the larger distance of $\simeq 519\, {\rm km}$. The linear
resolution in the finest grid, covering both \acp{NS} during the
inspiral and the merger remnant, is of ${\sim} 185\, {\rm m}$. We verify
the robustness of our results by repeating the 1.35+1.35 and 1.4+1.4
binary simulations (named after the isolation masses of the \acp{NS}) at
$50\%$ higher resolution. We also include neutrino cooling and lepton
number changes following \citet{radice:2016dwd}. As in
\citet{radice:2016dwd}, we use the finite-volume solver implemented in
\texttt{WhiskyTHC} with high-order reconstruction of the primitive
variables and an approximate Riemann solver. 

Ours are the first fully general-relativistic merger simulations
incorporating hyperons in a way consistent with all presently known
\ac{EOS} constraints.

\section{Results} %

Comparing the evolutions with the two \acp{EOS} we find, unsurprisingly,
negligible differences in the inspiral, because the \acp{EOS} agree at
$n \lesssim 2.5\, n_{\rm nuc}$. Even for the most massive
BHB$\Lambda\phi$-1.5+1.5 binary, the hyperon (mass) fraction remains
below $10\%$ during the inspiral.  This results in a small ${\sim} 5\%$
increase of the central density during the inspiral, it has a very
modest effect on the structure of the \acp{NS} and no measurable effect
on the \ac{GW} signal. This is in line with previous studies with
hyperons using other \acp{EOS} \citep{sekiguchi:2011mc} or analytic
approximations~\citep{chatziioannou:2015uea}.

The merger process is characterized by a sudden compression of the stars
followed by a rapid expansion; see the upper-right panel of
Fig.~\ref{fig:compactness} for the maximum density evolution of three
representative binaries. At this time, the production of
$\Lambda-$particles starts to become important for the dynamics. The
formation of hyperons in the interface layer between the \acp{NS}
during merger results in a catastrophic loss of pressure support, which
leads to a more violent merger. In the most extreme case,
BHB$\Lambda\phi$-1.5+1.5, this results in a temporary increase of the
maximum density from $\simeq 2\, n_{\rm nuc}$ to $\simeq 4.5\, n_{\rm
nuc}$ immediately at merger, followed by a violent centrifugal bounce
(Fig.~\ref{fig:compactness}).

After merger, the BHB$\Lambda\phi$ remnants are characterized by a
progressive increase of the hyperon fraction in their cores, which
causes their rapid contraction, while the DD2 remnants remain more
extended.  The central densities reached in the BHB$\Lambda\phi$
binaries correspond to isolated \ac{NS} masses of $1.8\, M_\odot - 2.0\,
M_\odot$. This contraction is also reflected in an increase in magnitude
of the binding energy of the binary (lower-right panel of
Fig.~\ref{fig:compactness}), which is offset by correspondingly larger
\ac{GW} luminosities. This holds until \ac{BH} formation, at which point
the \ac{GW} emission shuts off. This occurs at $12.0\, {\rm{ms}}$ and
$2.3\, {\rm{ms}}$ after merger for the most massive models with
hyperons, BHB$\Lambda\phi$-1.4+1.4 and BHB$\Lambda\phi$-1.5+1.5,
respectively. All other binaries result in remnants stable for the
entire simulation time.

These qualitative features of the dynamics are reflected in the \ac{GW}
strain, which we show, for the same three binaries, in
Fig.~\ref{fig:waves.spec}. As anticipated, the waveforms start to be
distinguishable only after merger with the BHB$\Lambda\phi$ binaries
becoming significantly louder in \acp{GW} after merger and until \ac{BH}
formation (if it occurs). The spectral content of the signals is shown
in the lower panels of Fig.~\ref{fig:waves.spec}. Although the
BHB$\Lambda\phi$ signals show significant excess power compared with the
DD2 ones, their peak frequencies are very similar. Indeed, the $f_2$
frequencies, which we extract from the spectrum of the entire postmerger
signal, show differences smaller than the scatter of the relations found
by \citet{bauswein:2011tp}, \citet{hotokezaka:2013iia}, and
\citet{bernuzzi:2014kca}. For most binaries, these differences are below
the nominal uncertainty of the Fourier transform ($\lesssim 50\, {\rm
Hz}$). Exceptions are the 1.4+1.4 and 1.5+1.5 binaries, where there are
signatures of early \ac{BH} formation. The former has $\Delta f_2 \simeq
250\, {\rm Hz}$, which is, however, still within the uncertainty of the
relations of \citet{bauswein:2011tp}, \citet{hotokezaka:2013iia}, and
\citet{bernuzzi:2015rla} for $f_2$. In the latter case, no post-merger
frequency can be extracted for the BHB$\Lambda\phi$ \ac{EOS}, due to the
prompt collapse.

We find evidence for small temporal drifts of the \ac{GW} peak
frequencies for both the DD2 and BHB$\Lambda\phi$ binaries, which
accelerate in the last few milliseconds prior to collapse. The
peak-frequency drift for the binaries with hyperons is comparable in
magnitude to that of the DD2 binaries and of other nucleonic \acp{EOS}
presented in the literature \citep{hotokezaka:2013iia, dietrich:2016hky,
rezzolla:2016nxn}. Our results do not seem to support the suggestion by
\citet{sekiguchi:2011mc} that the production of hyperons might be
encoded in the frequency evolution of the \ac{GW} signal. They reinforce
previous indications that the peak frequency of the \ac{GW} signal,
which has been the focus of all previous studies, is most sensitive to
the relatively low-density, low-temperature part of the \ac{EOS}
relevant for the inspiral \citep{bernuzzi:2015rla}.

It is important to remark that the BHB$\Lambda\phi$ waveforms are not
only louder than the DD2 waveform, they also have different amplitude
modulation and phase evolution. These differences make the
BHB$\Lambda\phi$ and DD2 waveforms distinguishable. To quantify this
observation, we window the waveforms to the interval $-1\, {\rm ms} \leq
t - t_{\rm mrg} \leq 20\, {\rm ms}$ and compute optimal \acp{SNR}, \ie,
assuming a perfect template, and \acp{FF} \citep{sathyaprakash:2009xs,
delpozzo:2014cla} for Adv.~LIGO, in its zero-detuning high-power
configuration \citep{ligo-sens-2010}, and for the \ac{ET}, in its ``D''
configuration \citep{punturo:2010zza, hild:2010id}. The \acp{FF} are
computed by maximizing the match between the BHB$\Lambda\phi$ and DD2
waveforms over time and phase shifts. In doing so, we implicitly assume
that the maximum match parameters between the two waveforms are the same
and that they coincide with their true values, which is reasonable since
these parameters could be extracted from the inspiral signal. In our
analysis, we also assume optimal orientation and sky position and limit
ourselves to the single detector case. Additionally, we estimate the
contribution of amplitude modulation by recomputing the \acp{FF} after
having stretched the waveforms to remove any difference in the
instantaneous phase evolution. We take the difference between the two
\acp{FF} as a conservative measure of uncertainty.

\begin{figure}
  \includegraphics[width=0.48\textwidth]{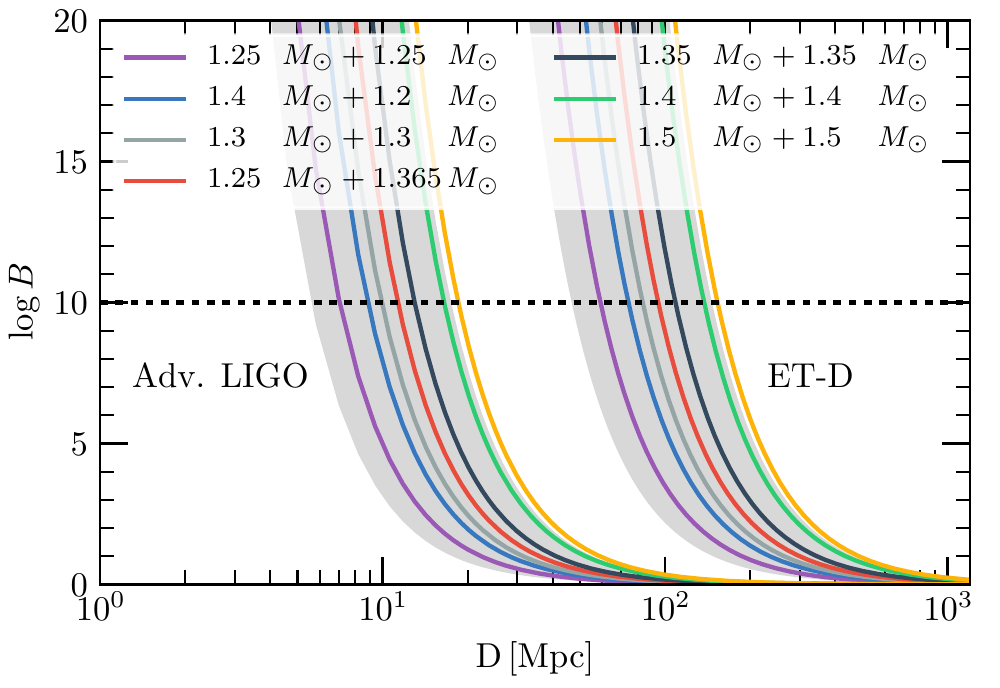}
  \caption{Distinguishability, as measured using the logarithm of the
  Bayes factor, between the DD2 and the BHB$\Lambda\phi$ \ac{EOS} with
  Adv.~LIGO and ET as a function of distance. Each line refers to a
  different binary. The grey shaded region denotes the uncertainty
  associated to the 1.35+1.35 binary, which is the most uncertain.}
  \label{fig:logBayes}
\end{figure}

Using \acp{FF} and \acp{SNR}, we estimate the logarithm of the Bayes'
factor against the presence of hyperons assuming DD2 to be Nature's true
\ac{EOS} following the approach proposed by \citet{vallisneri:2012qq}
and \citet{delpozzo:2014cla}. For each binary, we estimate the Bayes'
factor as a function of the distance. A similar calculation can be
repeated, with quantitatively very similar results, assuming
BHB$\Lambda\phi$ to be the true \ac{EOS} and computing the detectability
of hyperons. The results of the former analysis are shown in
Fig.~\ref{fig:logBayes}. Large values of $\log B$ indicate that strong
evidence against the presence of hyperons in the postmerger remnant
would be available. Using Jeffrey's scale, $\log B \geq 6$ would
constitute strong evidence and $\log B \geq 10$, would constitute a
decisive evidence \citep{kass:1995a}.  Correspondingly, in this
idealized scenario with two \ac{EOS} to discern, Adv.~LIGO could rule
out one of the two possibilities with a single merger at a distance of
up to ${\sim} 20\, {\rm Mpc}$, depending on the total mass of the
binary. This increases up to ${\sim} 200\, {\rm Mpc}$ with \ac{ET}. We
remark that these effective distances do not account for calibration
uncertainty of the detectors and non-optimal orientation of the
binaries, which would make these measurements even more challenging. At
the same time, we do not consider the possibility of stacking multiple
signals and/or data from multiple detectors, which could improve the
prospects for detection.

\section{Discussion} %

Our results show that the behavior of matter in the high density
postmerger stage is directly imprinted on the amplitude and phase of the
GW signal. The postmerger GW peak frequency is, instead, mainly
sensitive to matter properties at NS densities during inspiral. We have
demonstrated that \ac{GW} observations of \ac{NS} merger remnants can be
used to probe the \ac{EOS} of nuclear matter at extreme densities.

Here, we have only considered the possible presence of interacting
$\Lambda$-hyperons. In addition to $\Lambda$-hyperons, kaon or pion
condensates \citep{pons:2000iy}, transitions to quark matter
\citep{steiner:2000bi}, or the presence of other hyperons at high
densities are all possible.  Generally, phase transitions and extra
degrees of freedom soften the \ac{EOS} and increase the binding
energy of remnants with fixed baryon number, similar to the
BHB$\Lambda\phi$ model considered here. Therefore, we expect the
presence of other exotic phases of matter to have impacts qualitatively
similar to that of $\Lambda$-hyperons on the GW signal.

Our findings do not invalidate previously-proposed approaches to the
postmerger \ac{GW} data analysis \citep[\eg,][]{clark:2014wua,
clark:2015zxa}. We also do not exclude the possibility that constraints
on the high-density \ac{EOS} could be extracted from a more careful
analysis of the \ac{GW} peak frequency, going beyond existing
phenomenological fits, for example with the method suggested by
\citet{bauswein:2014qla}. However, on the basis of our results, we
advocate a more ambitious approach based on a Bayesian analysis of
\ac{GW} data using full waveform templates, with amplitude and phase
information, to extract the most likely values of parameters describing
the high-density \ac{EOS}. 

The kind of model-dependent inference we are proposing, which we have
shown to be able to probe the \ac{EOS} at the highest densities, will
require the availability of large banks of waveform templates. Possible
avenues to follow are either the construction of reduced order models
\citep{field:2013cfa, purrer:2014fza} or the use of a Gaussian process
regression strategy \citep{gair:2015nga, *Moore:2015sza}. Either
requires large databases of numerical waveforms covering the binary
parameter space for a variety of \acp{EOS}.

To be able to distinguish physical effects resulting in postmerger
waveform mismatch $M = 1-\mathrm{FF}$, it will be necessary to develop
template waveforms with a mismatch to the real signal $M_{\rm NR} \ll
M$. This accuracy requirement is stringent for low mass binaries, for
which the mismatches due to the appearance of hyperons are relatively
small. Nevertheless, they are within reach. For example, for the 1.4+1.4
binary, our high-resolution and standard-resolution data already have a
mismatch a factor ${\sim}2$ smaller than that due to hyperons. On the
other hand, systematic uncertainties due to missing physics still need
to be addressed. This will be the aim of our future work.

\begin{acknowledgments}
\acknowledgments
We thank S.~Hild for the ET-D noise curve data and the anonymous referee
for useful comments, and we acknowledge useful discussions with
A.~Burrows, T.~Dietrich, S.~Marka, L.~Rezzolla, B.~S.~Sathyaprakash, and
K.~Takami. DR gratefully acknowledges support from the Schmidt
Fellowship, the Sherman Fairchild Foundation and the
Max-Planck/Princeton Center (MPPC) for Plasma Physics (NSF PHY-1523261).
CDO was partially supported by NSF awards CAREER PHY-1151197,
PHY-1404569, and TCAN AST-1333520, and by the Sherman Fairchild
Foundation. The simulations were performed, on NSF XSEDE (TG-PHY160025),
and on NSF/NCSA Blue Waters (NSF PRAC ACI-1440083).
\end{acknowledgments}

\bibliography{references}

\begin{thebibliography}{50}%
\makeatletter
\providecommand \@ifxundefined [1]{%
 \@ifx{#1\undefined}
}%
\providecommand \@ifnum [1]{%
 \ifnum #1\expandafter \@firstoftwo
 \else \expandafter \@secondoftwo
 \fi
}%
\providecommand \@ifx [1]{%
 \ifx #1\expandafter \@firstoftwo
 \else \expandafter \@secondoftwo
 \fi
}%
\providecommand \natexlab [1]{#1}%
\providecommand \enquote  [1]{``#1''}%
\providecommand \bibnamefont  [1]{#1}%
\providecommand \bibfnamefont [1]{#1}%
\providecommand \citenamefont [1]{#1}%
\providecommand \href@noop [0]{\@secondoftwo}%
\providecommand \href [0]{\begingroup \@sanitize@url \@href}%
\providecommand \@href[1]{\@@startlink{#1}\@@href}%
\providecommand \@@href[1]{\endgroup#1\@@endlink}%
\providecommand \@sanitize@url [0]{\catcode `\\12\catcode `\$12\catcode
  `\&12\catcode `\#12\catcode `\^12\catcode `\_12\catcode `\%12\relax}%
\providecommand \@@startlink[1]{}%
\providecommand \@@endlink[0]{}%
\providecommand \url  [0]{\begingroup\@sanitize@url \@url }%
\providecommand \@url [1]{\endgroup\@href {#1}{\urlprefix }}%
\providecommand \urlprefix  [0]{URL }%
\providecommand \Eprint [0]{\href }%
\providecommand \doibase [0]{http://dx.doi.org/}%
\providecommand \selectlanguage [0]{\@gobble}%
\providecommand \bibinfo  [0]{\@secondoftwo}%
\providecommand \bibfield  [0]{\@secondoftwo}%
\providecommand \translation [1]{[#1]}%
\providecommand \BibitemOpen [0]{}%
\providecommand \bibitemStop [0]{}%
\providecommand \bibitemNoStop [0]{.\EOS\space}%
\providecommand \EOS [0]{\spacefactor3000\relax}%
\providecommand \BibitemShut  [1]{\csname bibitem#1\endcsname}%
\let\auto@bib@innerbib\@empty
\bibitem [{\citenamefont {Damour}\ \emph {et~al.}(2012)\citenamefont {Damour},
  \citenamefont {Nagar},\ and\ \citenamefont {Villain}}]{damour:2012yf}%
  \BibitemOpen
  \bibfield  {author} {\bibinfo {author} {\bibfnamefont {T.}~\bibnamefont
  {Damour}}, \bibinfo {author} {\bibfnamefont {A.}~\bibnamefont {Nagar}}, \
  and\ \bibinfo {author} {\bibfnamefont {L.}~\bibnamefont {Villain}},\ }\href
  {\doibase 10.1103/PhysRevD.85.123007} {\bibfield  {journal} {\bibinfo
  {journal} {Phys. Rev.}\ }\textbf {\bibinfo {volume} {D85}},\ \bibinfo {pages}
  {123007} (\bibinfo {year} {2012})},\ \Eprint {http://arxiv.org/abs/1203.4352}
  {arXiv:1203.4352 [gr-qc]} \BibitemShut {NoStop}%
\bibitem [{\citenamefont {Read}\ \emph {et~al.}(2013)\citenamefont {Read},
  \citenamefont {Baiotti}, \citenamefont {Creighton}, \citenamefont {Friedman},
  \citenamefont {Giacomazzo}, \citenamefont {Kyutoku}, \citenamefont
  {Markakis}, \citenamefont {Rezzolla}, \citenamefont {Shibata},\ and\
  \citenamefont {Taniguchi}}]{read:2013zra}%
  \BibitemOpen
  \bibfield  {author} {\bibinfo {author} {\bibfnamefont {J.~S.}\ \bibnamefont
  {Read}}, \bibinfo {author} {\bibfnamefont {L.}~\bibnamefont {Baiotti}},
  \bibinfo {author} {\bibfnamefont {J.~D.~E.}\ \bibnamefont {Creighton}},
  \bibinfo {author} {\bibfnamefont {J.~L.}\ \bibnamefont {Friedman}}, \bibinfo
  {author} {\bibfnamefont {B.}~\bibnamefont {Giacomazzo}}, \bibinfo {author}
  {\bibfnamefont {K.}~\bibnamefont {Kyutoku}}, \bibinfo {author} {\bibfnamefont
  {C.}~\bibnamefont {Markakis}}, \bibinfo {author} {\bibfnamefont
  {L.}~\bibnamefont {Rezzolla}}, \bibinfo {author} {\bibfnamefont
  {M.}~\bibnamefont {Shibata}}, \ and\ \bibinfo {author} {\bibfnamefont
  {K.}~\bibnamefont {Taniguchi}},\ }\href {\doibase 10.1103/PhysRevD.88.044042}
  {\bibfield  {journal} {\bibinfo  {journal} {Phys. Rev.}\ }\textbf {\bibinfo
  {volume} {D88}},\ \bibinfo {pages} {044042} (\bibinfo {year} {2013})},\
  \Eprint {http://arxiv.org/abs/1306.4065} {arXiv:1306.4065 [gr-qc]}
  \BibitemShut {NoStop}%
\bibitem [{\citenamefont {Del~Pozzo}\ \emph {et~al.}(2013)\citenamefont
  {Del~Pozzo}, \citenamefont {Li}, \citenamefont {Agathos}, \citenamefont {Van
  Den~Broeck},\ and\ \citenamefont {Vitale}}]{delpozzo:2013ala}%
  \BibitemOpen
  \bibfield  {author} {\bibinfo {author} {\bibfnamefont {W.}~\bibnamefont
  {Del~Pozzo}}, \bibinfo {author} {\bibfnamefont {T.~G.~F.}\ \bibnamefont
  {Li}}, \bibinfo {author} {\bibfnamefont {M.}~\bibnamefont {Agathos}},
  \bibinfo {author} {\bibfnamefont {C.}~\bibnamefont {Van Den~Broeck}}, \ and\
  \bibinfo {author} {\bibfnamefont {S.}~\bibnamefont {Vitale}},\ }\href
  {\doibase 10.1103/PhysRevLett.111.071101} {\bibfield  {journal} {\bibinfo
  {journal} {Phys. Rev. Lett.}\ }\textbf {\bibinfo {volume} {111}},\ \bibinfo
  {pages} {071101} (\bibinfo {year} {2013})},\ \Eprint
  {http://arxiv.org/abs/1307.8338} {arXiv:1307.8338 [gr-qc]} \BibitemShut
  {NoStop}%
\bibitem [{\citenamefont {Bernuzzi}\ \emph
  {et~al.}(2015{\natexlab{a}})\citenamefont {Bernuzzi}, \citenamefont {Nagar},
  \citenamefont {Dietrich},\ and\ \citenamefont {Damour}}]{bernuzzi:2014owa}%
  \BibitemOpen
  \bibfield  {author} {\bibinfo {author} {\bibfnamefont {S.}~\bibnamefont
  {Bernuzzi}}, \bibinfo {author} {\bibfnamefont {A.}~\bibnamefont {Nagar}},
  \bibinfo {author} {\bibfnamefont {T.}~\bibnamefont {Dietrich}}, \ and\
  \bibinfo {author} {\bibfnamefont {T.}~\bibnamefont {Damour}},\ }\href
  {\doibase 10.1103/PhysRevLett.114.161103} {\bibfield  {journal} {\bibinfo
  {journal} {Phys. Rev. Lett.}\ }\textbf {\bibinfo {volume} {114}},\ \bibinfo
  {pages} {161103} (\bibinfo {year} {2015}{\natexlab{a}})},\ \Eprint
  {http://arxiv.org/abs/1412.4553} {arXiv:1412.4553 [gr-qc]} \BibitemShut
  {NoStop}%
\bibitem [{\citenamefont {Hotokezaka}\ \emph {et~al.}(2016)\citenamefont
  {Hotokezaka}, \citenamefont {Kyutoku}, \citenamefont {Sekiguchi},\ and\
  \citenamefont {Shibata}}]{hotokezaka:2016bzh}%
  \BibitemOpen
  \bibfield  {author} {\bibinfo {author} {\bibfnamefont {K.}~\bibnamefont
  {Hotokezaka}}, \bibinfo {author} {\bibfnamefont {K.}~\bibnamefont {Kyutoku}},
  \bibinfo {author} {\bibfnamefont {Y.-i.}\ \bibnamefont {Sekiguchi}}, \ and\
  \bibinfo {author} {\bibfnamefont {M.}~\bibnamefont {Shibata}},\ }\href
  {\doibase 10.1103/PhysRevD.93.064082} {\bibfield  {journal} {\bibinfo
  {journal} {Phys. Rev.}\ }\textbf {\bibinfo {volume} {D93}},\ \bibinfo {pages}
  {064082} (\bibinfo {year} {2016})},\ \Eprint
  {http://arxiv.org/abs/1603.01286} {arXiv:1603.01286 [gr-qc]} \BibitemShut
  {NoStop}%
\bibitem [{\citenamefont {Hinderer}\ \emph {et~al.}(2016)\citenamefont
  {Hinderer} \emph {et~al.}}]{hinderer:2016eia}%
  \BibitemOpen
  \bibfield  {author} {\bibinfo {author} {\bibfnamefont {T.}~\bibnamefont
  {Hinderer}} \emph {et~al.},\ }\href {\doibase 10.1103/PhysRevLett.116.181101}
  {\bibfield  {journal} {\bibinfo  {journal} {Phys. Rev. Lett.}\ }\textbf
  {\bibinfo {volume} {116}},\ \bibinfo {pages} {181101} (\bibinfo {year}
  {2016})},\ \Eprint {http://arxiv.org/abs/1602.00599} {arXiv:1602.00599
  [gr-qc]} \BibitemShut {NoStop}%
\bibitem [{\citenamefont {Lackey}\ \emph {et~al.}(2016)\citenamefont {Lackey},
  \citenamefont {Bernuzzi}, \citenamefont {Galley}, \citenamefont {Meidam},\
  and\ \citenamefont {Van Den~Broeck}}]{lackey:2016krb}%
  \BibitemOpen
  \bibfield  {author} {\bibinfo {author} {\bibfnamefont {B.~D.}\ \bibnamefont
  {Lackey}}, \bibinfo {author} {\bibfnamefont {S.}~\bibnamefont {Bernuzzi}},
  \bibinfo {author} {\bibfnamefont {C.~R.}\ \bibnamefont {Galley}}, \bibinfo
  {author} {\bibfnamefont {J.}~\bibnamefont {Meidam}}, \ and\ \bibinfo {author}
  {\bibfnamefont {C.}~\bibnamefont {Van Den~Broeck}},\ }\href@noop {} {\
  (\bibinfo {year} {2016})},\ \Eprint {http://arxiv.org/abs/1610.04742}
  {arXiv:1610.04742 [gr-qc]} \BibitemShut {NoStop}%
\bibitem [{\citenamefont {Lattimer}(2012)}]{lattimer:2012nd}%
  \BibitemOpen
  \bibfield  {author} {\bibinfo {author} {\bibfnamefont {J.~M.}\ \bibnamefont
  {Lattimer}},\ }\href {\doibase 10.1146/annurev-nucl-102711-095018} {\bibfield
   {journal} {\bibinfo  {journal} {Ann. Rev. Nucl. Part. Sci.}\ }\textbf
  {\bibinfo {volume} {62}},\ \bibinfo {pages} {485} (\bibinfo {year} {2012})},\
  \Eprint {http://arxiv.org/abs/1305.3510} {arXiv:1305.3510 [nucl-th]}
  \BibitemShut {NoStop}%
\bibitem [{\citenamefont {Demorest}\ \emph {et~al.}(2010)\citenamefont
  {Demorest}, \citenamefont {Pennucci}, \citenamefont {Ransom}, \citenamefont
  {Roberts},\ and\ \citenamefont {Hessels}}]{demorest:2010bx}%
  \BibitemOpen
  \bibfield  {author} {\bibinfo {author} {\bibfnamefont {P.}~\bibnamefont
  {Demorest}}, \bibinfo {author} {\bibfnamefont {T.}~\bibnamefont {Pennucci}},
  \bibinfo {author} {\bibfnamefont {S.}~\bibnamefont {Ransom}}, \bibinfo
  {author} {\bibfnamefont {M.}~\bibnamefont {Roberts}}, \ and\ \bibinfo
  {author} {\bibfnamefont {J.}~\bibnamefont {Hessels}},\ }\href {\doibase
  10.1038/nature09466} {\bibfield  {journal} {\bibinfo  {journal} {Nature}\
  }\textbf {\bibinfo {volume} {467}},\ \bibinfo {pages} {1081} (\bibinfo {year}
  {2010})},\ \Eprint {http://arxiv.org/abs/1010.5788} {arXiv:1010.5788
  [astro-ph.HE]} \BibitemShut {NoStop}%
\bibitem [{\citenamefont {Antoniadis}\ \emph {et~al.}(2013)\citenamefont
  {Antoniadis} \emph {et~al.}}]{antoniadis:2013pzd}%
  \BibitemOpen
  \bibfield  {author} {\bibinfo {author} {\bibfnamefont {J.}~\bibnamefont
  {Antoniadis}} \emph {et~al.},\ }\href {\doibase 10.1126/science.1233232}
  {\bibfield  {journal} {\bibinfo  {journal} {Science}\ }\textbf {\bibinfo
  {volume} {340}},\ \bibinfo {pages} {6131} (\bibinfo {year} {2013})},\ \Eprint
  {http://arxiv.org/abs/1304.6875} {arXiv:1304.6875 [astro-ph.HE]} \BibitemShut
  {NoStop}%
\bibitem [{\citenamefont {Rosswog}\ and\ \citenamefont
  {Davies}(2003)}]{rosswog:2001fh}%
  \BibitemOpen
  \bibfield  {author} {\bibinfo {author} {\bibfnamefont {S.}~\bibnamefont
  {Rosswog}}\ and\ \bibinfo {author} {\bibfnamefont {M.~B.}\ \bibnamefont
  {Davies}},\ }\href {\doibase 10.1046/j.1365-2966.2003.07032.x} {\bibfield
  {journal} {\bibinfo  {journal} {Mon. Not. Roy. Astron. Soc.}\ }\textbf
  {\bibinfo {volume} {345}},\ \bibinfo {pages} {1077} (\bibinfo {year}
  {2003})},\ \Eprint {http://arxiv.org/abs/astro-ph/0110180}
  {arXiv:astro-ph/0110180 [astro-ph]} \BibitemShut {NoStop}%
\bibitem [{\citenamefont {Shibata}\ and\ \citenamefont
  {Taniguchi}(2006)}]{shibata:2006nm}%
  \BibitemOpen
  \bibfield  {author} {\bibinfo {author} {\bibfnamefont {M.}~\bibnamefont
  {Shibata}}\ and\ \bibinfo {author} {\bibfnamefont {K.}~\bibnamefont
  {Taniguchi}},\ }\href {\doibase 10.1103/PhysRevD.73.064027} {\bibfield
  {journal} {\bibinfo  {journal} {Phys. Rev.}\ }\textbf {\bibinfo {volume}
  {D73}},\ \bibinfo {pages} {064027} (\bibinfo {year} {2006})},\ \Eprint
  {http://arxiv.org/abs/astro-ph/0603145} {arXiv:astro-ph/0603145 [astro-ph]}
  \BibitemShut {NoStop}%
\bibitem [{\citenamefont {Baiotti}\ \emph {et~al.}(2008)\citenamefont
  {Baiotti}, \citenamefont {Giacomazzo},\ and\ \citenamefont
  {Rezzolla}}]{baiotti:2008ra}%
  \BibitemOpen
  \bibfield  {author} {\bibinfo {author} {\bibfnamefont {L.}~\bibnamefont
  {Baiotti}}, \bibinfo {author} {\bibfnamefont {B.}~\bibnamefont {Giacomazzo}},
  \ and\ \bibinfo {author} {\bibfnamefont {L.}~\bibnamefont {Rezzolla}},\
  }\href {\doibase 10.1103/PhysRevD.78.084033} {\bibfield  {journal} {\bibinfo
  {journal} {Phys. Rev.}\ }\textbf {\bibinfo {volume} {D78}},\ \bibinfo {pages}
  {084033} (\bibinfo {year} {2008})},\ \Eprint {http://arxiv.org/abs/0804.0594}
  {arXiv:0804.0594 [gr-qc]} \BibitemShut {NoStop}%
\bibitem [{\citenamefont {Sekiguchi}\ \emph
  {et~al.}(2011{\natexlab{a}})\citenamefont {Sekiguchi}, \citenamefont
  {Kiuchi}, \citenamefont {Kyutoku},\ and\ \citenamefont
  {Shibata}}]{sekiguchi:2011zd}%
  \BibitemOpen
  \bibfield  {author} {\bibinfo {author} {\bibfnamefont {Y.}~\bibnamefont
  {Sekiguchi}}, \bibinfo {author} {\bibfnamefont {K.}~\bibnamefont {Kiuchi}},
  \bibinfo {author} {\bibfnamefont {K.}~\bibnamefont {Kyutoku}}, \ and\
  \bibinfo {author} {\bibfnamefont {M.}~\bibnamefont {Shibata}},\ }\href
  {\doibase 10.1103/PhysRevLett.107.051102} {\bibfield  {journal} {\bibinfo
  {journal} {Phys. Rev. Lett.}\ }\textbf {\bibinfo {volume} {107}},\ \bibinfo
  {pages} {051102} (\bibinfo {year} {2011}{\natexlab{a}})},\ \Eprint
  {http://arxiv.org/abs/1105.2125} {arXiv:1105.2125 [gr-qc]} \BibitemShut
  {NoStop}%
\bibitem [{\citenamefont {Palenzuela}\ \emph {et~al.}(2015)\citenamefont
  {Palenzuela}, \citenamefont {Liebling}, \citenamefont {Neilsen},
  \citenamefont {Lehner}, \citenamefont {Caballero}, \citenamefont {O'Connor},\
  and\ \citenamefont {Anderson}}]{palenzuela:2015dqa}%
  \BibitemOpen
  \bibfield  {author} {\bibinfo {author} {\bibfnamefont {C.}~\bibnamefont
  {Palenzuela}}, \bibinfo {author} {\bibfnamefont {S.~L.}\ \bibnamefont
  {Liebling}}, \bibinfo {author} {\bibfnamefont {D.}~\bibnamefont {Neilsen}},
  \bibinfo {author} {\bibfnamefont {L.}~\bibnamefont {Lehner}}, \bibinfo
  {author} {\bibfnamefont {O.~L.}\ \bibnamefont {Caballero}}, \bibinfo {author}
  {\bibfnamefont {E.}~\bibnamefont {O'Connor}}, \ and\ \bibinfo {author}
  {\bibfnamefont {M.}~\bibnamefont {Anderson}},\ }\href {\doibase
  10.1103/PhysRevD.92.044045} {\bibfield  {journal} {\bibinfo  {journal} {Phys.
  Rev.}\ }\textbf {\bibinfo {volume} {D92}},\ \bibinfo {pages} {044045}
  (\bibinfo {year} {2015})},\ \Eprint {http://arxiv.org/abs/1505.01607}
  {arXiv:1505.01607 [gr-qc]} \BibitemShut {NoStop}%
\bibitem [{\citenamefont {Foucart}\ \emph {et~al.}(2016)\citenamefont
  {Foucart}, \citenamefont {Haas}, \citenamefont {Duez}, \citenamefont
  {O'Connor}, \citenamefont {Ott}, \citenamefont {Roberts}, \citenamefont
  {Kidder}, \citenamefont {Lippuner}, \citenamefont {Pfeiffer},\ and\
  \citenamefont {Scheel}}]{foucart:2015gaa}%
  \BibitemOpen
  \bibfield  {author} {\bibinfo {author} {\bibfnamefont {F.}~\bibnamefont
  {Foucart}}, \bibinfo {author} {\bibfnamefont {R.}~\bibnamefont {Haas}},
  \bibinfo {author} {\bibfnamefont {M.~D.}\ \bibnamefont {Duez}}, \bibinfo
  {author} {\bibfnamefont {E.}~\bibnamefont {O'Connor}}, \bibinfo {author}
  {\bibfnamefont {C.~D.}\ \bibnamefont {Ott}}, \bibinfo {author} {\bibfnamefont
  {L.}~\bibnamefont {Roberts}}, \bibinfo {author} {\bibfnamefont {L.~E.}\
  \bibnamefont {Kidder}}, \bibinfo {author} {\bibfnamefont {J.}~\bibnamefont
  {Lippuner}}, \bibinfo {author} {\bibfnamefont {H.~P.}\ \bibnamefont
  {Pfeiffer}}, \ and\ \bibinfo {author} {\bibfnamefont {M.~A.}\ \bibnamefont
  {Scheel}},\ }\href {\doibase 10.1103/PhysRevD.93.044019} {\bibfield
  {journal} {\bibinfo  {journal} {Phys. Rev.}\ }\textbf {\bibinfo {volume}
  {D93}},\ \bibinfo {pages} {044019} (\bibinfo {year} {2016})},\ \Eprint
  {http://arxiv.org/abs/1510.06398} {arXiv:1510.06398 [astro-ph.HE]}
  \BibitemShut {NoStop}%
\bibitem [{\citenamefont {Baiotti}\ and\ \citenamefont
  {Rezzolla}(2016)}]{baiotti:2016qnr}%
  \BibitemOpen
  \bibfield  {author} {\bibinfo {author} {\bibfnamefont {L.}~\bibnamefont
  {Baiotti}}\ and\ \bibinfo {author} {\bibfnamefont {L.}~\bibnamefont
  {Rezzolla}},\ }\href@noop {} {\  (\bibinfo {year} {2016})},\ \Eprint
  {http://arxiv.org/abs/1607.03540} {arXiv:1607.03540 [gr-qc]} \BibitemShut
  {NoStop}%
\bibitem [{\citenamefont {Bernuzzi}\ \emph
  {et~al.}(2015{\natexlab{b}})\citenamefont {Bernuzzi}, \citenamefont {Radice},
  \citenamefont {Ott}, \citenamefont {Roberts}, \citenamefont {Moesta},\ and\
  \citenamefont {Galeazzi}}]{bernuzzi:2015opx}%
  \BibitemOpen
  \bibfield  {author} {\bibinfo {author} {\bibfnamefont {S.}~\bibnamefont
  {Bernuzzi}}, \bibinfo {author} {\bibfnamefont {D.}~\bibnamefont {Radice}},
  \bibinfo {author} {\bibfnamefont {C.~D.}\ \bibnamefont {Ott}}, \bibinfo
  {author} {\bibfnamefont {L.~F.}\ \bibnamefont {Roberts}}, \bibinfo {author}
  {\bibfnamefont {P.}~\bibnamefont {Moesta}}, \ and\ \bibinfo {author}
  {\bibfnamefont {F.}~\bibnamefont {Galeazzi}},\ }\href@noop {} {\  (\bibinfo
  {year} {2015}{\natexlab{b}})},\ \Eprint {http://arxiv.org/abs/1512.06397}
  {arXiv:1512.06397 [gr-qc]} \BibitemShut {NoStop}%
\bibitem [{\citenamefont {Bauswein}\ and\ \citenamefont
  {Janka}(2012)}]{bauswein:2011tp}%
  \BibitemOpen
  \bibfield  {author} {\bibinfo {author} {\bibfnamefont {A.}~\bibnamefont
  {Bauswein}}\ and\ \bibinfo {author} {\bibfnamefont {H.~T.}\ \bibnamefont
  {Janka}},\ }\href {\doibase 10.1103/PhysRevLett.108.011101} {\bibfield
  {journal} {\bibinfo  {journal} {Phys. Rev. Lett.}\ }\textbf {\bibinfo
  {volume} {108}},\ \bibinfo {pages} {011101} (\bibinfo {year} {2012})},\
  \Eprint {http://arxiv.org/abs/1106.1616} {arXiv:1106.1616 [astro-ph.SR]}
  \BibitemShut {NoStop}%
\bibitem [{\citenamefont {Stergioulas}\ \emph {et~al.}(2011)\citenamefont
  {Stergioulas}, \citenamefont {Bauswein}, \citenamefont {Zagkouris},\ and\
  \citenamefont {Janka}}]{stergioulas:2011gd}%
  \BibitemOpen
  \bibfield  {author} {\bibinfo {author} {\bibfnamefont {N.}~\bibnamefont
  {Stergioulas}}, \bibinfo {author} {\bibfnamefont {A.}~\bibnamefont
  {Bauswein}}, \bibinfo {author} {\bibfnamefont {K.}~\bibnamefont {Zagkouris}},
  \ and\ \bibinfo {author} {\bibfnamefont {H.-T.}\ \bibnamefont {Janka}},\
  }\href {\doibase 10.1111/j.1365-2966.2011.19493.x} {\bibfield  {journal}
  {\bibinfo  {journal} {Mon. Not. Roy. Astron. Soc.}\ }\textbf {\bibinfo
  {volume} {418}},\ \bibinfo {pages} {427} (\bibinfo {year} {2011})},\ \Eprint
  {http://arxiv.org/abs/1105.0368} {arXiv:1105.0368 [gr-qc]} \BibitemShut
  {NoStop}%
\bibitem [{\citenamefont {Takami}\ \emph {et~al.}(2014)\citenamefont {Takami},
  \citenamefont {Rezzolla},\ and\ \citenamefont {Baiotti}}]{takami:2014zpa}%
  \BibitemOpen
  \bibfield  {author} {\bibinfo {author} {\bibfnamefont {K.}~\bibnamefont
  {Takami}}, \bibinfo {author} {\bibfnamefont {L.}~\bibnamefont {Rezzolla}}, \
  and\ \bibinfo {author} {\bibfnamefont {L.}~\bibnamefont {Baiotti}},\ }\href
  {\doibase 10.1103/PhysRevLett.113.091104} {\bibfield  {journal} {\bibinfo
  {journal} {Phys. Rev. Lett.}\ }\textbf {\bibinfo {volume} {113}},\ \bibinfo
  {pages} {091104} (\bibinfo {year} {2014})},\ \Eprint
  {http://arxiv.org/abs/1403.5672} {arXiv:1403.5672 [gr-qc]} \BibitemShut
  {NoStop}%
\bibitem [{\citenamefont {Bauswein}\ and\ \citenamefont
  {Stergioulas}(2015)}]{bauswein:2015yca}%
  \BibitemOpen
  \bibfield  {author} {\bibinfo {author} {\bibfnamefont {A.}~\bibnamefont
  {Bauswein}}\ and\ \bibinfo {author} {\bibfnamefont {N.}~\bibnamefont
  {Stergioulas}},\ }\href {\doibase 10.1103/PhysRevD.91.124056} {\bibfield
  {journal} {\bibinfo  {journal} {Phys. Rev.}\ }\textbf {\bibinfo {volume}
  {D91}},\ \bibinfo {pages} {124056} (\bibinfo {year} {2015})},\ \Eprint
  {http://arxiv.org/abs/1502.03176} {arXiv:1502.03176 [astro-ph.SR]}
  \BibitemShut {NoStop}%
\bibitem [{\citenamefont {Rezzolla}\ and\ \citenamefont
  {Takami}(2016)}]{rezzolla:2016nxn}%
  \BibitemOpen
  \bibfield  {author} {\bibinfo {author} {\bibfnamefont {L.}~\bibnamefont
  {Rezzolla}}\ and\ \bibinfo {author} {\bibfnamefont {K.}~\bibnamefont
  {Takami}},\ }\href {\doibase 10.1103/PhysRevD.93.124051} {\bibfield
  {journal} {\bibinfo  {journal} {Phys. Rev.}\ }\textbf {\bibinfo {volume}
  {D93}},\ \bibinfo {pages} {124051} (\bibinfo {year} {2016})},\ \Eprint
  {http://arxiv.org/abs/1604.00246} {arXiv:1604.00246 [gr-qc]} \BibitemShut
  {NoStop}%
\bibitem [{\citenamefont {Dietrich}\ \emph {et~al.}(2016)\citenamefont
  {Dietrich}, \citenamefont {Ujevic}, \citenamefont {Tichy}, \citenamefont
  {Bernuzzi},\ and\ \citenamefont {Bruegmann}}]{dietrich:2016hky}%
  \BibitemOpen
  \bibfield  {author} {\bibinfo {author} {\bibfnamefont {T.}~\bibnamefont
  {Dietrich}}, \bibinfo {author} {\bibfnamefont {M.}~\bibnamefont {Ujevic}},
  \bibinfo {author} {\bibfnamefont {W.}~\bibnamefont {Tichy}}, \bibinfo
  {author} {\bibfnamefont {S.}~\bibnamefont {Bernuzzi}}, \ and\ \bibinfo
  {author} {\bibfnamefont {B.}~\bibnamefont {Bruegmann}},\ }\href@noop {} {\
  (\bibinfo {year} {2016})},\ \Eprint {http://arxiv.org/abs/1607.06636}
  {arXiv:1607.06636 [gr-qc]} \BibitemShut {NoStop}%
\bibitem [{\citenamefont {Hotokezaka}\ \emph {et~al.}(2013)\citenamefont
  {Hotokezaka}, \citenamefont {Kiuchi}, \citenamefont {Kyutoku}, \citenamefont
  {Muranushi}, \citenamefont {Sekiguchi}, \citenamefont {Shibata},\ and\
  \citenamefont {Taniguchi}}]{hotokezaka:2013iia}%
  \BibitemOpen
  \bibfield  {author} {\bibinfo {author} {\bibfnamefont {K.}~\bibnamefont
  {Hotokezaka}}, \bibinfo {author} {\bibfnamefont {K.}~\bibnamefont {Kiuchi}},
  \bibinfo {author} {\bibfnamefont {K.}~\bibnamefont {Kyutoku}}, \bibinfo
  {author} {\bibfnamefont {T.}~\bibnamefont {Muranushi}}, \bibinfo {author}
  {\bibfnamefont {Y.}~\bibnamefont {Sekiguchi}}, \bibinfo {author}
  {\bibfnamefont {M.}~\bibnamefont {Shibata}}, \ and\ \bibinfo {author}
  {\bibfnamefont {K.}~\bibnamefont {Taniguchi}},\ }\href {\doibase
  10.1103/PhysRevD.88.044026} {\bibfield  {journal} {\bibinfo  {journal} {Phys.
  Rev.}\ }\textbf {\bibinfo {volume} {D88}},\ \bibinfo {pages} {044026}
  (\bibinfo {year} {2013})},\ \Eprint {http://arxiv.org/abs/1307.5888}
  {arXiv:1307.5888 [astro-ph.HE]} \BibitemShut {NoStop}%
\bibitem [{\citenamefont {Bauswein}\ \emph {et~al.}(2014)\citenamefont
  {Bauswein}, \citenamefont {Stergioulas},\ and\ \citenamefont
  {Janka}}]{bauswein:2014qla}%
  \BibitemOpen
  \bibfield  {author} {\bibinfo {author} {\bibfnamefont {A.}~\bibnamefont
  {Bauswein}}, \bibinfo {author} {\bibfnamefont {N.}~\bibnamefont
  {Stergioulas}}, \ and\ \bibinfo {author} {\bibfnamefont {H.~T.}\ \bibnamefont
  {Janka}},\ }\href {\doibase 10.1103/PhysRevD.90.023002} {\bibfield  {journal}
  {\bibinfo  {journal} {Phys. Rev.}\ }\textbf {\bibinfo {volume} {D90}},\
  \bibinfo {pages} {023002} (\bibinfo {year} {2014})},\ \Eprint
  {http://arxiv.org/abs/1403.5301} {arXiv:1403.5301 [astro-ph.SR]} \BibitemShut
  {NoStop}%
\bibitem [{\citenamefont {Bernuzzi}\ \emph
  {et~al.}(2015{\natexlab{c}})\citenamefont {Bernuzzi}, \citenamefont
  {Dietrich},\ and\ \citenamefont {Nagar}}]{bernuzzi:2015rla}%
  \BibitemOpen
  \bibfield  {author} {\bibinfo {author} {\bibfnamefont {S.}~\bibnamefont
  {Bernuzzi}}, \bibinfo {author} {\bibfnamefont {T.}~\bibnamefont {Dietrich}},
  \ and\ \bibinfo {author} {\bibfnamefont {A.}~\bibnamefont {Nagar}},\ }\href
  {\doibase 10.1103/PhysRevLett.115.091101} {\bibfield  {journal} {\bibinfo
  {journal} {Phys. Rev. Lett.}\ }\textbf {\bibinfo {volume} {115}},\ \bibinfo
  {pages} {091101} (\bibinfo {year} {2015}{\natexlab{c}})},\ \Eprint
  {http://arxiv.org/abs/1504.01764} {arXiv:1504.01764 [gr-qc]} \BibitemShut
  {NoStop}%
\bibitem [{\citenamefont {Bernuzzi}\ \emph {et~al.}(2014)\citenamefont
  {Bernuzzi}, \citenamefont {Nagar}, \citenamefont {Balmelli}, \citenamefont
  {Dietrich},\ and\ \citenamefont {Ujevic}}]{bernuzzi:2014kca}%
  \BibitemOpen
  \bibfield  {author} {\bibinfo {author} {\bibfnamefont {S.}~\bibnamefont
  {Bernuzzi}}, \bibinfo {author} {\bibfnamefont {A.}~\bibnamefont {Nagar}},
  \bibinfo {author} {\bibfnamefont {S.}~\bibnamefont {Balmelli}}, \bibinfo
  {author} {\bibfnamefont {T.}~\bibnamefont {Dietrich}}, \ and\ \bibinfo
  {author} {\bibfnamefont {M.}~\bibnamefont {Ujevic}},\ }\href {\doibase
  10.1103/PhysRevLett.112.201101} {\bibfield  {journal} {\bibinfo  {journal}
  {Phys. Rev. Lett.}\ }\textbf {\bibinfo {volume} {112}},\ \bibinfo {pages}
  {201101} (\bibinfo {year} {2014})},\ \Eprint {http://arxiv.org/abs/1402.6244}
  {arXiv:1402.6244 [gr-qc]} \BibitemShut {NoStop}%
\bibitem [{\citenamefont {Typel}\ \emph {et~al.}(2010)\citenamefont {Typel},
  \citenamefont {Ropke}, \citenamefont {Klahn}, \citenamefont {Blaschke},\ and\
  \citenamefont {Wolter}}]{typel:2009sy}%
  \BibitemOpen
  \bibfield  {author} {\bibinfo {author} {\bibfnamefont {S.}~\bibnamefont
  {Typel}}, \bibinfo {author} {\bibfnamefont {G.}~\bibnamefont {Ropke}},
  \bibinfo {author} {\bibfnamefont {T.}~\bibnamefont {Klahn}}, \bibinfo
  {author} {\bibfnamefont {D.}~\bibnamefont {Blaschke}}, \ and\ \bibinfo
  {author} {\bibfnamefont {H.~H.}\ \bibnamefont {Wolter}},\ }\href {\doibase
  10.1103/PhysRevC.81.015803} {\bibfield  {journal} {\bibinfo  {journal} {Phys.
  Rev.}\ }\textbf {\bibinfo {volume} {C81}},\ \bibinfo {pages} {015803}
  (\bibinfo {year} {2010})},\ \Eprint {http://arxiv.org/abs/0908.2344}
  {arXiv:0908.2344 [nucl-th]} \BibitemShut {NoStop}%
\bibitem [{\citenamefont {Hempel}\ and\ \citenamefont
  {Schaffner-Bielich}(2010)}]{hempel:2009mc}%
  \BibitemOpen
  \bibfield  {author} {\bibinfo {author} {\bibfnamefont {M.}~\bibnamefont
  {Hempel}}\ and\ \bibinfo {author} {\bibfnamefont {J.}~\bibnamefont
  {Schaffner-Bielich}},\ }\href {\doibase 10.1016/j.nuclphysa.2010.02.010}
  {\bibfield  {journal} {\bibinfo  {journal} {Nucl. Phys.}\ }\textbf {\bibinfo
  {volume} {A837}},\ \bibinfo {pages} {210} (\bibinfo {year} {2010})},\ \Eprint
  {http://arxiv.org/abs/0911.4073} {arXiv:0911.4073 [nucl-th]} \BibitemShut
  {NoStop}%
\bibitem [{\citenamefont {Banik}\ \emph {et~al.}(2014)\citenamefont {Banik},
  \citenamefont {Hempel},\ and\ \citenamefont {Bandyopadhyay}}]{banik:2014qja}%
  \BibitemOpen
  \bibfield  {author} {\bibinfo {author} {\bibfnamefont {S.}~\bibnamefont
  {Banik}}, \bibinfo {author} {\bibfnamefont {M.}~\bibnamefont {Hempel}}, \
  and\ \bibinfo {author} {\bibfnamefont {D.}~\bibnamefont {Bandyopadhyay}},\
  }\href {\doibase 10.1088/0067-0049/214/2/22} {\bibfield  {journal} {\bibinfo
  {journal} {Astrophys. J. Suppl.}\ }\textbf {\bibinfo {volume} {214}},\
  \bibinfo {pages} {22} (\bibinfo {year} {2014})},\ \Eprint
  {http://arxiv.org/abs/1404.6173} {arXiv:1404.6173 [astro-ph.HE]} \BibitemShut
  {NoStop}%
\bibitem [{\citenamefont {Radice}\ \emph {et~al.}(2014)\citenamefont {Radice},
  \citenamefont {Rezzolla},\ and\ \citenamefont {Galeazzi}}]{radice:2013xpa}%
  \BibitemOpen
  \bibfield  {author} {\bibinfo {author} {\bibfnamefont {D.}~\bibnamefont
  {Radice}}, \bibinfo {author} {\bibfnamefont {L.}~\bibnamefont {Rezzolla}}, \
  and\ \bibinfo {author} {\bibfnamefont {F.}~\bibnamefont {Galeazzi}},\ }\href
  {\doibase 10.1088/0264-9381/31/7/075012} {\bibfield  {journal} {\bibinfo
  {journal} {Class. Quant. Grav.}\ }\textbf {\bibinfo {volume} {31}},\ \bibinfo
  {pages} {075012} (\bibinfo {year} {2014})},\ \Eprint
  {http://arxiv.org/abs/1312.5004} {arXiv:1312.5004 [gr-qc]} \BibitemShut
  {NoStop}%
\bibitem [{\citenamefont {Radice}\ \emph {et~al.}(2016)\citenamefont {Radice},
  \citenamefont {Galeazzi}, \citenamefont {Lippuner}, \citenamefont {Roberts},
  \citenamefont {Ott},\ and\ \citenamefont {Rezzolla}}]{radice:2016dwd}%
  \BibitemOpen
  \bibfield  {author} {\bibinfo {author} {\bibfnamefont {D.}~\bibnamefont
  {Radice}}, \bibinfo {author} {\bibfnamefont {F.}~\bibnamefont {Galeazzi}},
  \bibinfo {author} {\bibfnamefont {J.}~\bibnamefont {Lippuner}}, \bibinfo
  {author} {\bibfnamefont {L.~F.}\ \bibnamefont {Roberts}}, \bibinfo {author}
  {\bibfnamefont {C.~D.}\ \bibnamefont {Ott}}, \ and\ \bibinfo {author}
  {\bibfnamefont {L.}~\bibnamefont {Rezzolla}},\ }\href {\doibase
  10.1093/mnras/stw1227} {\bibfield  {journal} {\bibinfo  {journal} {Mon. Not.
  Roy. Astron. Soc.}\ }\textbf {\bibinfo {volume} {460}},\ \bibinfo {pages}
  {3255} (\bibinfo {year} {2016})},\ \Eprint {http://arxiv.org/abs/1601.02426}
  {arXiv:1601.02426 [astro-ph.HE]} \BibitemShut {NoStop}%
\bibitem [{\citenamefont {Sekiguchi}\ \emph
  {et~al.}(2011{\natexlab{b}})\citenamefont {Sekiguchi}, \citenamefont
  {Kiuchi}, \citenamefont {Kyutoku},\ and\ \citenamefont
  {Shibata}}]{sekiguchi:2011mc}%
  \BibitemOpen
  \bibfield  {author} {\bibinfo {author} {\bibfnamefont {Y.}~\bibnamefont
  {Sekiguchi}}, \bibinfo {author} {\bibfnamefont {K.}~\bibnamefont {Kiuchi}},
  \bibinfo {author} {\bibfnamefont {K.}~\bibnamefont {Kyutoku}}, \ and\
  \bibinfo {author} {\bibfnamefont {M.}~\bibnamefont {Shibata}},\ }\href
  {\doibase 10.1103/PhysRevLett.107.211101} {\bibfield  {journal} {\bibinfo
  {journal} {Phys. Rev. Lett.}\ }\textbf {\bibinfo {volume} {107}},\ \bibinfo
  {pages} {211101} (\bibinfo {year} {2011}{\natexlab{b}})},\ \Eprint
  {http://arxiv.org/abs/1110.4442} {arXiv:1110.4442 [astro-ph.HE]} \BibitemShut
  {NoStop}%
\bibitem [{\citenamefont {Chatziioannou}\ \emph {et~al.}(2015)\citenamefont
  {Chatziioannou}, \citenamefont {Yagi}, \citenamefont {Klein}, \citenamefont
  {Cornish},\ and\ \citenamefont {Yunes}}]{chatziioannou:2015uea}%
  \BibitemOpen
  \bibfield  {author} {\bibinfo {author} {\bibfnamefont {K.}~\bibnamefont
  {Chatziioannou}}, \bibinfo {author} {\bibfnamefont {K.}~\bibnamefont {Yagi}},
  \bibinfo {author} {\bibfnamefont {A.}~\bibnamefont {Klein}}, \bibinfo
  {author} {\bibfnamefont {N.}~\bibnamefont {Cornish}}, \ and\ \bibinfo
  {author} {\bibfnamefont {N.}~\bibnamefont {Yunes}},\ }\href {\doibase
  10.1103/PhysRevD.92.104008} {\bibfield  {journal} {\bibinfo  {journal} {Phys.
  Rev.}\ }\textbf {\bibinfo {volume} {D92}},\ \bibinfo {pages} {104008}
  (\bibinfo {year} {2015})},\ \Eprint {http://arxiv.org/abs/1508.02062}
  {arXiv:1508.02062 [gr-qc]} \BibitemShut {NoStop}%
\bibitem [{\citenamefont {Sathyaprakash}\ and\ \citenamefont
  {Schutz}(2009)}]{sathyaprakash:2009xs}%
  \BibitemOpen
  \bibfield  {author} {\bibinfo {author} {\bibfnamefont {B.~S.}\ \bibnamefont
  {Sathyaprakash}}\ and\ \bibinfo {author} {\bibfnamefont {B.~F.}\ \bibnamefont
  {Schutz}},\ }\href {\doibase 10.12942/lrr-2009-2} {\bibfield  {journal}
  {\bibinfo  {journal} {Living Rev. Rel.}\ }\textbf {\bibinfo {volume} {12}},\
  \bibinfo {pages} {2} (\bibinfo {year} {2009})},\ \Eprint
  {http://arxiv.org/abs/0903.0338} {arXiv:0903.0338 [gr-qc]} \BibitemShut
  {NoStop}%
\bibitem [{\citenamefont {Del~Pozzo}\ \emph {et~al.}(2014)\citenamefont
  {Del~Pozzo}, \citenamefont {Grover}, \citenamefont {Mandel},\ and\
  \citenamefont {Vecchio}}]{delpozzo:2014cla}%
  \BibitemOpen
  \bibfield  {author} {\bibinfo {author} {\bibfnamefont {W.}~\bibnamefont
  {Del~Pozzo}}, \bibinfo {author} {\bibfnamefont {K.}~\bibnamefont {Grover}},
  \bibinfo {author} {\bibfnamefont {I.}~\bibnamefont {Mandel}}, \ and\ \bibinfo
  {author} {\bibfnamefont {A.}~\bibnamefont {Vecchio}},\ }\href {\doibase
  10.1088/0264-9381/31/20/205006} {\bibfield  {journal} {\bibinfo  {journal}
  {Class. Quant. Grav.}\ }\textbf {\bibinfo {volume} {31}},\ \bibinfo {pages}
  {205006} (\bibinfo {year} {2014})},\ \Eprint {http://arxiv.org/abs/1408.2356}
  {arXiv:1408.2356 [gr-qc]} \BibitemShut {NoStop}%
\bibitem [{\citenamefont {Shoemaker}(2010)}]{ligo-sens-2010}%
  \BibitemOpen
  \bibfield  {author} {\bibinfo {author} {\bibfnamefont {D.}~\bibnamefont
  {Shoemaker}},\ }\href
  {https://dcc.ligo.org/cgi-bin/DocDB/ShowDocument?docid=t0900288} {\emph
  {\bibinfo {title} {Advanced LIGO anticipated sensitivity curves}}},\ \bibinfo
  {type} {Tech. Rep.}\ \bibinfo {number} {LIGO-T0900288-v3}\ (\bibinfo
  {institution} {LIGO Scientific Collaboration},\ \bibinfo {year}
  {2010})\BibitemShut {NoStop}%
\bibitem [{\citenamefont {Punturo}\ \emph {et~al.}(2010)\citenamefont {Punturo}
  \emph {et~al.}}]{punturo:2010zza}%
  \BibitemOpen
  \bibfield  {author} {\bibinfo {author} {\bibfnamefont {M.}~\bibnamefont
  {Punturo}} \emph {et~al.},\ }\bibfield  {booktitle} {\emph {\bibinfo
  {booktitle} {{Gravitational waves. Proceedings, 8th Edoardo Amaldi
  Conference, Amaldi 8, New York, USA, June 22-26, 2009}}},\ }\href {\doibase
  10.1088/0264-9381/27/8/084007} {\bibfield  {journal} {\bibinfo  {journal}
  {Class. Quant. Grav.}\ }\textbf {\bibinfo {volume} {27}},\ \bibinfo {pages}
  {084007} (\bibinfo {year} {2010})}\BibitemShut {NoStop}%
\bibitem [{\citenamefont {Hild}\ \emph {et~al.}(2011)\citenamefont {Hild} \emph
  {et~al.}}]{hild:2010id}%
  \BibitemOpen
  \bibfield  {author} {\bibinfo {author} {\bibfnamefont {S.}~\bibnamefont
  {Hild}} \emph {et~al.},\ }\href {\doibase 10.1088/0264-9381/28/9/094013}
  {\bibfield  {journal} {\bibinfo  {journal} {Class. Quant. Grav.}\ }\textbf
  {\bibinfo {volume} {28}},\ \bibinfo {pages} {094013} (\bibinfo {year}
  {2011})},\ \Eprint {http://arxiv.org/abs/1012.0908} {arXiv:1012.0908 [gr-qc]}
  \BibitemShut {NoStop}%
\bibitem [{\citenamefont {Vallisneri}(2012)}]{vallisneri:2012qq}%
  \BibitemOpen
  \bibfield  {author} {\bibinfo {author} {\bibfnamefont {M.}~\bibnamefont
  {Vallisneri}},\ }\href {\doibase 10.1103/PhysRevD.86.082001} {\bibfield
  {journal} {\bibinfo  {journal} {Phys. Rev.}\ }\textbf {\bibinfo {volume}
  {D86}},\ \bibinfo {pages} {082001} (\bibinfo {year} {2012})},\ \Eprint
  {http://arxiv.org/abs/1207.4759} {arXiv:1207.4759 [gr-qc]} \BibitemShut
  {NoStop}%
\bibitem [{\citenamefont {Kass}\ and\ \citenamefont
  {Raftery}(1995)}]{kass:1995a}%
  \BibitemOpen
  \bibfield  {author} {\bibinfo {author} {\bibfnamefont {R.~E.}\ \bibnamefont
  {Kass}}\ and\ \bibinfo {author} {\bibfnamefont {A.~E.}\ \bibnamefont
  {Raftery}},\ }\href {\doibase 10.1080/01621459.1995.10476572} {\bibfield
  {journal} {\bibinfo  {journal} {Journal of the American Statistical
  Association}\ }\textbf {\bibinfo {volume} {90}},\ \bibinfo {pages} {773}
  (\bibinfo {year} {1995})}\BibitemShut {NoStop}%
\bibitem [{\citenamefont {Pons}\ \emph {et~al.}(2000)\citenamefont {Pons},
  \citenamefont {Reddy}, \citenamefont {Ellis}, \citenamefont {Prakash},\ and\
  \citenamefont {Lattimer}}]{pons:2000iy}%
  \BibitemOpen
  \bibfield  {author} {\bibinfo {author} {\bibfnamefont {J.~A.}\ \bibnamefont
  {Pons}}, \bibinfo {author} {\bibfnamefont {S.}~\bibnamefont {Reddy}},
  \bibinfo {author} {\bibfnamefont {P.~J.}\ \bibnamefont {Ellis}}, \bibinfo
  {author} {\bibfnamefont {M.}~\bibnamefont {Prakash}}, \ and\ \bibinfo
  {author} {\bibfnamefont {J.~M.}\ \bibnamefont {Lattimer}},\ }\href {\doibase
  10.1103/PhysRevC.62.035803} {\bibfield  {journal} {\bibinfo  {journal} {Phys.
  Rev.}\ }\textbf {\bibinfo {volume} {C62}},\ \bibinfo {pages} {035803}
  (\bibinfo {year} {2000})},\ \Eprint {http://arxiv.org/abs/nucl-th/0003008}
  {arXiv:nucl-th/0003008 [nucl-th]} \BibitemShut {NoStop}%
\bibitem [{\citenamefont {Steiner}\ \emph {et~al.}(2000)\citenamefont
  {Steiner}, \citenamefont {Prakash},\ and\ \citenamefont
  {Lattimer}}]{steiner:2000bi}%
  \BibitemOpen
  \bibfield  {author} {\bibinfo {author} {\bibfnamefont {A.}~\bibnamefont
  {Steiner}}, \bibinfo {author} {\bibfnamefont {M.}~\bibnamefont {Prakash}}, \
  and\ \bibinfo {author} {\bibfnamefont {J.~M.}\ \bibnamefont {Lattimer}},\
  }\href {\doibase 10.1016/S0370-2693(00)00780-2} {\bibfield  {journal}
  {\bibinfo  {journal} {Phys. Lett.}\ }\textbf {\bibinfo {volume} {B486}},\
  \bibinfo {pages} {239} (\bibinfo {year} {2000})},\ \Eprint
  {http://arxiv.org/abs/nucl-th/0003066} {arXiv:nucl-th/0003066 [nucl-th]}
  \BibitemShut {NoStop}%
\bibitem [{\citenamefont {Clark}\ \emph {et~al.}(2014)\citenamefont {Clark},
  \citenamefont {Bauswein}, \citenamefont {Cadonati}, \citenamefont {Janka},
  \citenamefont {Pankow},\ and\ \citenamefont {Stergioulas}}]{clark:2014wua}%
  \BibitemOpen
  \bibfield  {author} {\bibinfo {author} {\bibfnamefont {J.}~\bibnamefont
  {Clark}}, \bibinfo {author} {\bibfnamefont {A.}~\bibnamefont {Bauswein}},
  \bibinfo {author} {\bibfnamefont {L.}~\bibnamefont {Cadonati}}, \bibinfo
  {author} {\bibfnamefont {H.~T.}\ \bibnamefont {Janka}}, \bibinfo {author}
  {\bibfnamefont {C.}~\bibnamefont {Pankow}}, \ and\ \bibinfo {author}
  {\bibfnamefont {N.}~\bibnamefont {Stergioulas}},\ }\href {\doibase
  10.1103/PhysRevD.90.062004} {\bibfield  {journal} {\bibinfo  {journal} {Phys.
  Rev.}\ }\textbf {\bibinfo {volume} {D90}},\ \bibinfo {pages} {062004}
  (\bibinfo {year} {2014})},\ \Eprint {http://arxiv.org/abs/1406.5444}
  {arXiv:1406.5444 [astro-ph.HE]} \BibitemShut {NoStop}%
\bibitem [{\citenamefont {Clark}\ \emph {et~al.}(2016)\citenamefont {Clark},
  \citenamefont {Bauswein}, \citenamefont {Stergioulas},\ and\ \citenamefont
  {Shoemaker}}]{clark:2015zxa}%
  \BibitemOpen
  \bibfield  {author} {\bibinfo {author} {\bibfnamefont {J.~A.}\ \bibnamefont
  {Clark}}, \bibinfo {author} {\bibfnamefont {A.}~\bibnamefont {Bauswein}},
  \bibinfo {author} {\bibfnamefont {N.}~\bibnamefont {Stergioulas}}, \ and\
  \bibinfo {author} {\bibfnamefont {D.}~\bibnamefont {Shoemaker}},\ }\href
  {\doibase 10.1088/0264-9381/33/8/085003} {\bibfield  {journal} {\bibinfo
  {journal} {Class. Quant. Grav.}\ }\textbf {\bibinfo {volume} {33}},\ \bibinfo
  {pages} {085003} (\bibinfo {year} {2016})},\ \Eprint
  {http://arxiv.org/abs/1509.08522} {arXiv:1509.08522 [astro-ph.HE]}
  \BibitemShut {NoStop}%
\bibitem [{\citenamefont {Field}\ \emph {et~al.}(2014)\citenamefont {Field},
  \citenamefont {Galley}, \citenamefont {Hesthaven}, \citenamefont {Kaye},\
  and\ \citenamefont {Tiglio}}]{field:2013cfa}%
  \BibitemOpen
  \bibfield  {author} {\bibinfo {author} {\bibfnamefont {S.~E.}\ \bibnamefont
  {Field}}, \bibinfo {author} {\bibfnamefont {C.~R.}\ \bibnamefont {Galley}},
  \bibinfo {author} {\bibfnamefont {J.~S.}\ \bibnamefont {Hesthaven}}, \bibinfo
  {author} {\bibfnamefont {J.}~\bibnamefont {Kaye}}, \ and\ \bibinfo {author}
  {\bibfnamefont {M.}~\bibnamefont {Tiglio}},\ }\href {\doibase
  10.1103/PhysRevX.4.031006} {\bibfield  {journal} {\bibinfo  {journal} {Phys.
  Rev.}\ }\textbf {\bibinfo {volume} {X4}},\ \bibinfo {pages} {031006}
  (\bibinfo {year} {2014})},\ \Eprint {http://arxiv.org/abs/1308.3565}
  {arXiv:1308.3565 [gr-qc]} \BibitemShut {NoStop}%
\bibitem [{\citenamefont {Pürrer}(2014)}]{purrer:2014fza}%
  \BibitemOpen
  \bibfield  {author} {\bibinfo {author} {\bibfnamefont {M.}~\bibnamefont
  {Pürrer}},\ }\href {\doibase 10.1088/0264-9381/31/19/195010} {\bibfield
  {journal} {\bibinfo  {journal} {Class. Quant. Grav.}\ }\textbf {\bibinfo
  {volume} {31}},\ \bibinfo {pages} {195010} (\bibinfo {year} {2014})},\
  \Eprint {http://arxiv.org/abs/1402.4146} {arXiv:1402.4146 [gr-qc]}
  \BibitemShut {NoStop}%
\bibitem [{\citenamefont {Gair}\ and\ \citenamefont
  {Moore}(2015)}]{gair:2015nga}%
  \BibitemOpen
  \bibfield  {author} {\bibinfo {author} {\bibfnamefont {J.~R.}\ \bibnamefont
  {Gair}}\ and\ \bibinfo {author} {\bibfnamefont {C.~J.}\ \bibnamefont
  {Moore}},\ }\href {\doibase 10.1103/PhysRevD.91.124062} {\bibfield  {journal}
  {\bibinfo  {journal} {Phys. Rev.}\ }\textbf {\bibinfo {volume} {D91}},\
  \bibinfo {pages} {124062} (\bibinfo {year} {2015})},\ \Eprint
  {http://arxiv.org/abs/1504.02767} {arXiv:1504.02767 [gr-qc]} \BibitemShut
  {NoStop}%
\bibitem [{\citenamefont {Moore}\ \emph {et~al.}(2016)\citenamefont {Moore},
  \citenamefont {Berry}, \citenamefont {Chua},\ and\ \citenamefont
  {Gair}}]{Moore:2015sza}%
  \BibitemOpen
  \bibfield  {author} {\bibinfo {author} {\bibfnamefont {C.~J.}\ \bibnamefont
  {Moore}}, \bibinfo {author} {\bibfnamefont {C.~P.~L.}\ \bibnamefont {Berry}},
  \bibinfo {author} {\bibfnamefont {A.~J.~K.}\ \bibnamefont {Chua}}, \ and\
  \bibinfo {author} {\bibfnamefont {J.~R.}\ \bibnamefont {Gair}},\ }\href
  {\doibase 10.1103/PhysRevD.93.064001} {\bibfield  {journal} {\bibinfo
  {journal} {Phys. Rev.}\ }\textbf {\bibinfo {volume} {D93}},\ \bibinfo {pages}
  {064001} (\bibinfo {year} {2016})},\ \Eprint
  {http://arxiv.org/abs/1509.04066} {arXiv:1509.04066 [gr-qc]} \BibitemShut
  {NoStop}%
\end{thebibliography}

\acrodef{ADM}{Arnowitt-Deser-Misner}
\acrodef{AMR}{adaptive mesh-refinement}
\acrodef{BH}{black hole}
\acrodef{BBH}{binary black-hole}
\acrodef{BHNS}{black-hole neutron-star}
\acrodef{BNS}{binary neutron star}
\acrodef{CCSN}{core-collapse supernova}
\acrodefplural{CCSN}[CCSNe]{core-collapse supernovae}
\acrodef{CMA}{consistent multi-fluid advection}
\acrodef{DG}{discontinuous Galerkin}
\acrodef{HMNS}{hypermassive neutron star}
\acrodef{EM}{electromagnetic}
\acrodef{ET}{Einstein Telescope}
\acrodef{EOB}{effective-one-body}
\acrodef{EOS}{equation of state}
\acrodefplural{EOS}[EOS]{equations of state}
\acrodef{FF}{fitting factor}
\acrodef{GR}{general relativistic}
\acrodef{GRHD}{general-relativistic hydrodynamics}
\acrodef{GW}{gravitational wave}
\acrodef{LIA}{linear interaction analysis}
\acrodef{LES}{large-eddy simulation}
\acrodefplural{LES}[LES]{large-eddy simulations}
\acrodef{NR}{numerical relativity}
\acrodef{NS}{neutron star}
\acrodef{PNS}{protoneutron star}
\acrodef{SASI}{standing accretion shock instability}
\acrodef{SGRB}{short gamma-ray burst}
\acrodef{SN}{supernova}
\acrodefplural{SN}[SNe]{supernovae}
\acrodef{SNR}{signal-to-noise ratio}

\end{document}